\documentclass[preprint]{revtex4}
\usepackage{graphicx,color} 
\usepackage{bm}       
\usepackage{amsmath}  
\usepackage{amssymb}

\begin{document}
\title{Properties of Lithium-11 and Carbon-22 at leading order in halo effective field theory\footnote{Contribution to the $21^\text{st}$ International Conference on Few-Body Problems in Physics}}

\author{Bijaya~Acharya}
\email{bacharya@utk.edu}
\affiliation{Department of Physics and Astronomy, University of Tennessee, Knoxville, TN 37996, USA}
\affiliation{
           Institute of Nuclear and Particle Physics and Department of Physics and Astronomy, \\ Ohio University, Athens, OH 45701, USA 
          }

\author{Daniel~Phillips}
\email{phillips@phy.ohiou.edu}
\affiliation{
           Institute of Nuclear and Particle Physics and Department of Physics and Astronomy, \\ Ohio University, Athens, OH 45701, USA 
          }

\begin{abstract}

We study the $^{11}\mathrm{Li}$ and $^{22}\mathrm{C}$ nuclei at leading order (LO) in halo effective field theory (Halo EFT).
Using the value of the $^{22}\mathrm{C}$ rms matter radius deduced in Ref.~\cite{Tanaka:2010zza} as an input in a LO calculation, 
we simultaneously constrained the values of the two-neutron (2$n$) separation energy of $^{22}\mathrm{C}$ and the virtual-state energy of the 
$^{20}\mathrm{C}-$neutron system (hereafter denoted $^{21}$C). The 1$-\sigma$ uncertainty of the input rms matter radius datum, 
along with the theory error estimated from the anticipated size of the higher-order terms in the Halo EFT expansion, 
gave an upper bound of about 100~keV for the 2$n$ separation energy. 
We also study the electric dipole excitation of 2$n$ halo nuclei to a continuum state of two neutrons and the core at LO in Halo EFT. 
We first compare our results with the $^{11}\mathrm{Li}$ data from a Coulomb dissociation experiment and obtain good agreement
within the theoretical uncertainty of a LO calculation.
We then obtain the low-energy spectrum of $B(E1)$ of this transition at several different values of the 
2$n$ separation energy of $^{22}\mathrm{C}$ and the virtual-state energy of $^{21}\mathrm{C}$.
Our predictions can be compared to the outcome of an ongoing experiment 
on the Coulomb dissociation of $^{22}\mathrm{C}$ to obtain tighter constraints on the two- and three-body energies in the $^{22}\mathrm{C}$ system.

\end{abstract}

\smallskip

\maketitle

\section{Introduction}

The separation of scales between the size of the core and its distance from the halo nucleons allows the low-energy properites of halo nuclei to 
be studied using Halo EFT~\cite{Bertulani:2002sz,Bedaque:2003wa}, which is written in terms of the core and halo nucleons as degrees of freedom. 
Halo EFT yields relations for the low-energy observables as systematic expansions in the ratio of the short-distance scale set by the core size and excitation energies 
to the long-distance scale associated with the properties of the halo nucleons. At LO, the three-body wavefunction of 
the 2$n$ halo nucleus is constructed with zero-range two-body interactions, which can be 
completely characterized by the neutron-neutron ($nn$) and the neutron-core ($nc$) scattering lengths~\cite{Kaplan:1998we}. 
However, a three-body coupling also enters at LO~\cite{Bedaque:1998kg}, necessitating the use of one piece of three-body data 
as input to render the theory predictive. It is convenient to fix the three-body force by 
requiring the three-body bound state to lie at $-E_B$, where $E_B$ is the 2$n$ separation energy.
The only inputs to the equations that describe a 2$n$ halo are, therefore, $E_B$ together with the energies of the $nc$ virtual/real bound state, 
$E_{nc}$, and the $nn$ virtual bound state, $E_{nn}$. The effects of interactions that are higher order in the Halo EFT power counting are 
estimated from the size of the ignored higher-order terms and then included as theory error bands. 

In Ref.~\cite{Tanaka:2010zza},
Tanaka et al. measured the reaction cross-section of $^{22}\text{C}$ on a hydrogen target and, using Glauber calculations, deduced a $^{22}\text{C}$ 
rms matter radius of $5.4\pm 0.9$~fm, implying that $^{22}\text{C}$ is an S-wave two-neturon halo nucleus. 
This conclusion is also supported by data on high-energy two-neutron removal from ${}^{22}$C~\cite{Kobayashi:2011mm}. 
We used Halo EFT in Ref.~\cite{Acharya:2013aea}, to calculate the rms matter radius of $^{22}$C as a model-independent 
function of $E_B$ and $E_{nc}$. Since the  virtual-state energy of the unbound~\cite{Langevin} $^{21}\text{C}$ 
is not well known~\cite{Mosby:2013bix}, we used Halo EFT to find constraints in the $(E_B,E_{nc})$ plane using Tanaka et al.'s value of the rms matter radius.

We have also derived universal relations for the electric dipole excitation of two-neutron halo nuclei into the three-body continuum 
consisting of the core and the two neutrons in Halo EFT. Our LO calculation of the $B(E1)$ of this transition includes all possible rescatterings with 
S-wave $nn$ and $nc$ interactions, in both the initial and the final state. We compare our results with the $^{11}\mathrm{Li}$ data from 
Ref.~\cite{Nakamura:2006zz} and obtain a good agreement within the theoretical uncertainty. 
We predict the $B(E1)$ spectrum of $^{22}\mathrm{C}$ for selected values of $E_B$ and $E_{nc}$. 
These findings will be published in Ref.~\cite{Acharya:tobepublished}.
 
\section{Matter radius constraints on binding energy}

\begin{figure}[ht]
\begin{center}
\includegraphics[width=0.75\textwidth]{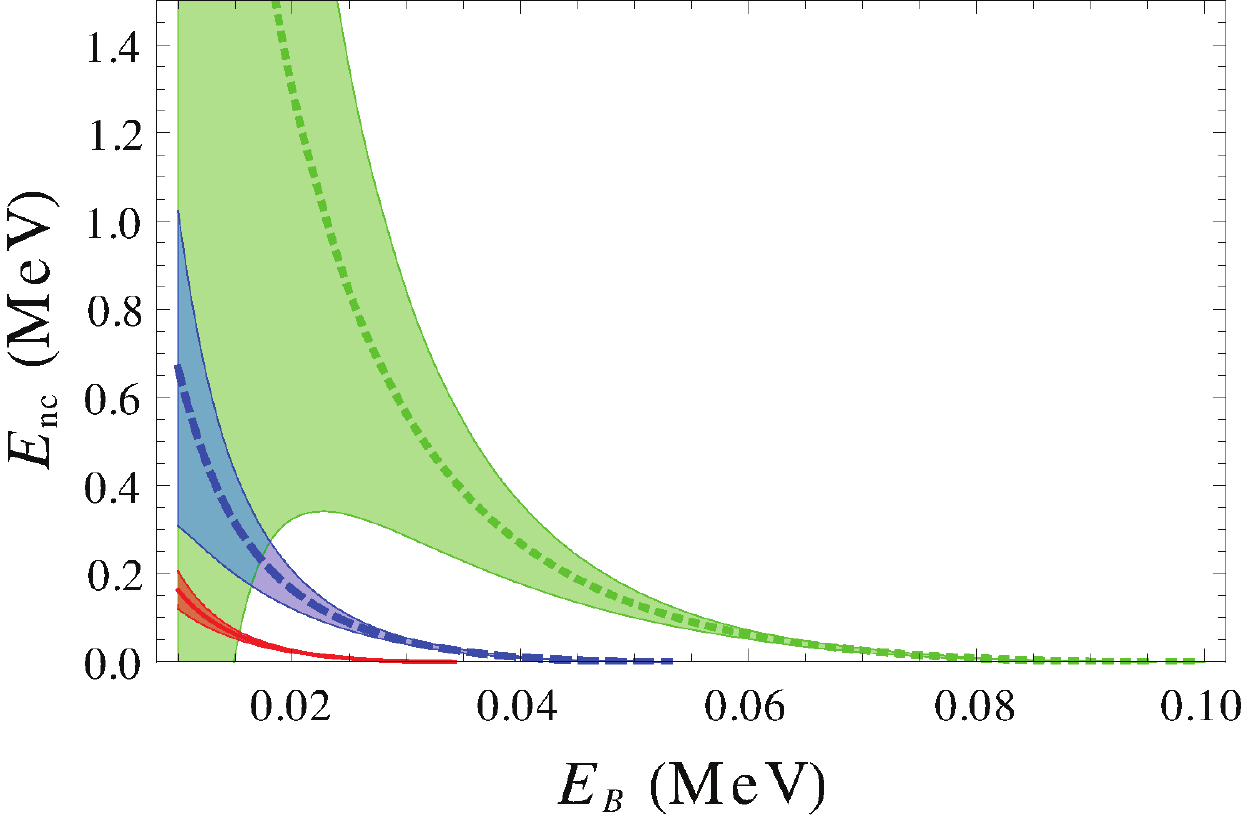}
\caption{Plots of $\sqrt{\langle R^2 \rangle} $~=~5.4~fm (blue, dashed), 6.3~fm (red, solid), and 4.5~fm (green, dotted), with their 
theoretical error bands, in the $(E_B,E_{nc})$ plane. (Published in 
Ref.~\cite{Acharya:2013aea}.)}
\label{fig:contourplots}
\end{center}
 \end{figure}

In Fig.~\ref{fig:contourplots}, we plot the sets of  ($E_B$, $E_{nc}$) values that give a $^{22}\mathrm{C}$ rms matter radius, $\sqrt{\langle R^2 \rangle}$, 
of 4.5~fm, 5.4~fm and 6.3~fm, along with the theoretical error bands.  All sets of $E_B$ and $E_{nc}$ values in the plotted region that lie within the area 
bounded by the edges of these bands  give an rms matter radius that is consistent with the value Tanaka et al. extracted within the combined ($1-\sigma$) experimental and theoretical errors. 
The figure shows that, regardless of the value of the $^{21}\text{C}$ virtual energy, 
Tanaka et al.'s experimental result puts a model-independent upper limit of 100~keV on the 2$n$ separation energy of $^{22}\text{C}$.
This is to be compared with another theoretical analysis of the matter radius datum of Ref.~\cite{Tanaka:2010zza} in a three-body model by Ref.~\cite{Yamashita:2011cb}, which set an upper bound of 
120~keV on $E_B$. Similarly, Ref.~\cite{Fortune:2012zzb} used a correlation between the binding energy and the matter radius derived from a potential model to 
exclude $E_B>220$~keV. Our constraint is stricter than the ones set by these studies. 
Although our conclusion is consistent with the experimental value of $-140~(460)$~keV from a direct mass measurement~\cite{Gaudefroy:2012qe}, 
more studies are needed to further reduce the large uncertainty in the 2$n$ separation energy. 
In this spirit, we study the $E1$ excitation of 2$n$ halo nuclei to the three-body continuum.

\section{The {\emph B}({\emph E}1) spectrum}

\begin{figure}[ht]
\centering
\includegraphics[width=0.75\textwidth]{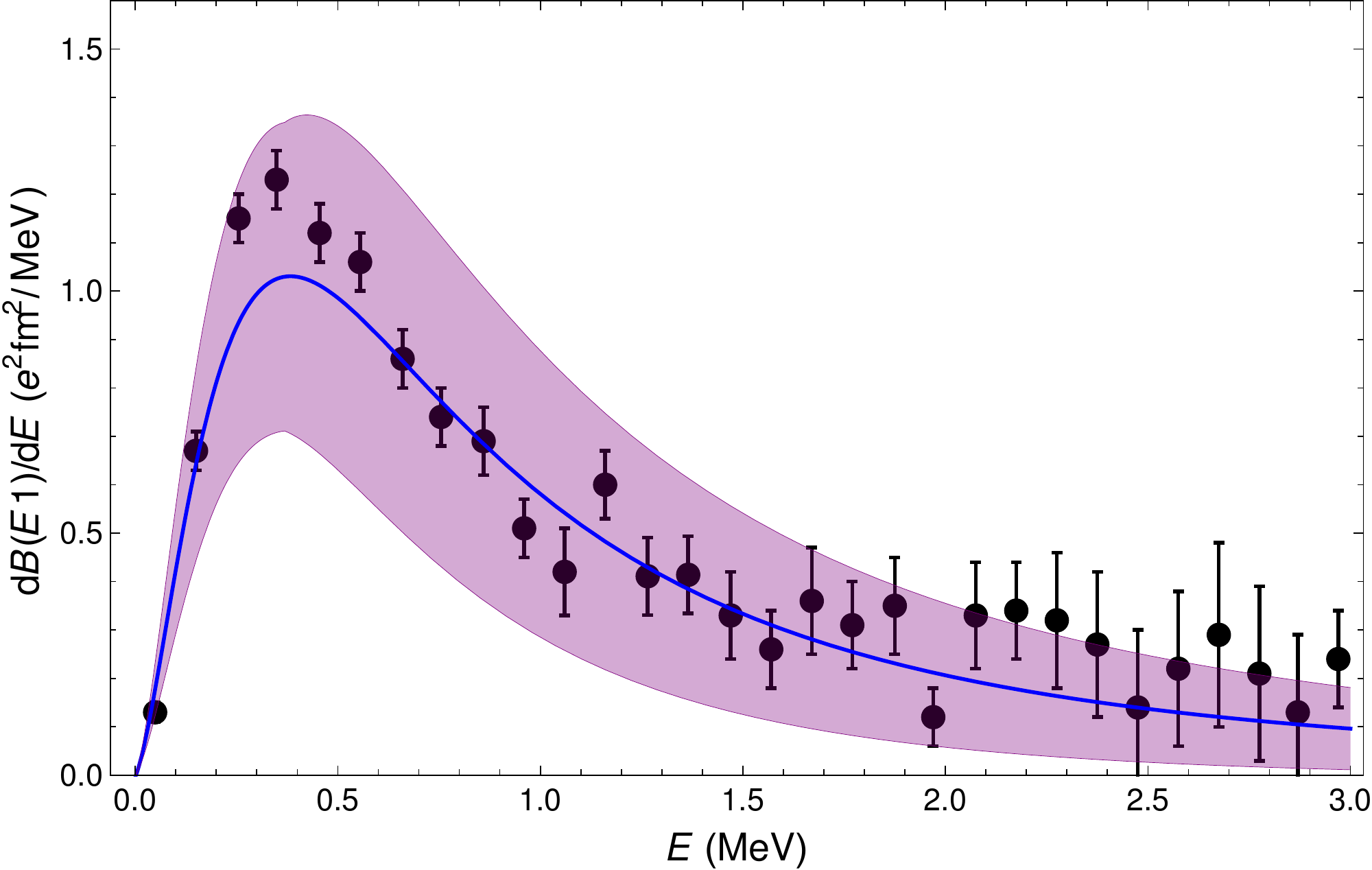}
\caption{The dipole response spectrum for $^{11}\mathrm{Li}$ after folding with the detector resolution (blue curve) 
with the theory error (purple band), and data from Ref.~\cite{Nakamura:2006zz}.} 
\label{fig:li11}
\end{figure}

We first present the result of our LO Halo EFT calculation of the $B(E1)$ for the break up of $^{11}\mathrm{Li}$ into 
$^{9}\mathrm{Li}$ and two neutrons at energy $E$ in their center of mass frame. 
Only S-wave $^9\mathrm{Li}-n$ interactions are included. 
After folding with the detector resolution, we obtain the curve shown in Fig.~\ref{fig:li11}
for $E_B=369.15(65)~\mathrm{keV}$~\cite{Smith:2008zh} and $E_{nc}=26~\mathrm{keV}$~\cite{NNDC}. 
The sensitivity to changes in $E_{nc}$ is much smaller than the EFT error, represented by the purple band.
Within the uncertainty of a LO calculation, a good agreement with the RIKEN data~\cite{Nakamura:2006zz} is seen, 
despite the fact that $^{10}\mathrm{Li}$ has a low-lying P-wave resonance which is not included in this calculation.

\begin{figure}[h]
\centering
\includegraphics[width=0.75\textwidth]{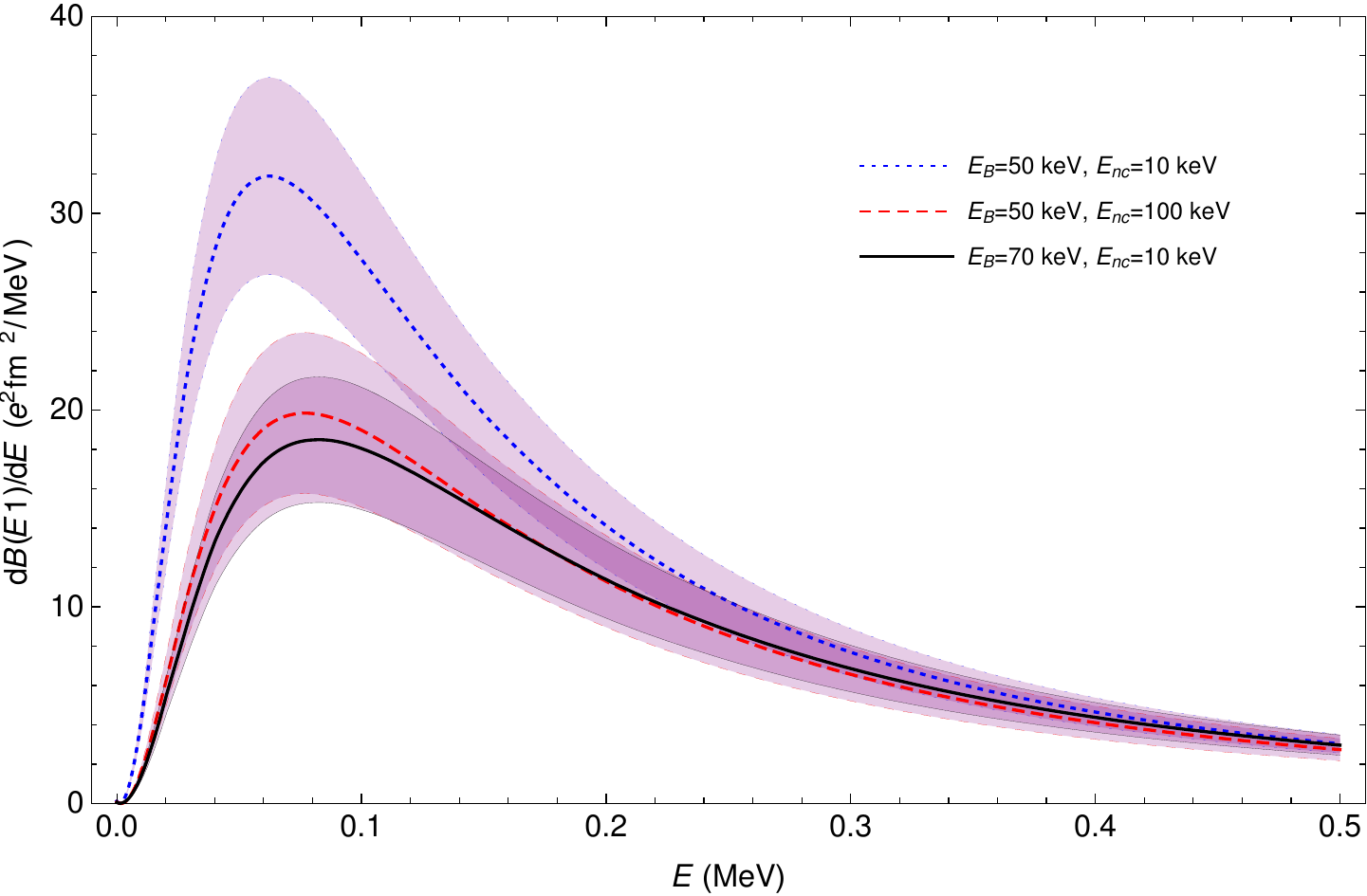}
\caption{The dipole response spectrum for $^{22}\mathrm{C}$ for $E_B=50$~keV,~$E_{nc}=10$~keV (blue, dotted); $E_B=50$~keV,~$E_{nc}=100$~keV (red, dashed) 
and $E_B=70$~keV,~$E_{nc}=10$~keV (black), with their EFT error bands.} 
\label{fig:c22}
\end{figure}

Figure~\ref{fig:c22} shows the dipole response spectrum for the break up of $^{22}\mathrm{C}$ into $^{20}\mathrm{C}$ and neutrons for three different combinations  
of $E_B$ and $E_{nc}$ which lie within the $1-\sigma$ confidence region shown in Fig.~\ref{fig:contourplots}. These results agree qualitatively with those of a 
potential model calculation by Ref.~\cite{Ershov:2012fy}. A comparison of Fig.~\ref{fig:c22} with the forthcoming data~\cite{Nakamura:2013conference} 
can provide further constraints on the $(E_B,E_{nc})$ plane. However, the individual values of these energies thus extracted will have large error bars because different sets of $(E_B,E_{nc})$ values 
can give similar curves. This ambiguity can be removed by looking at the neutron-momentum distribution of the Coulomb dissociation cross section~\cite{Acharya:tobepublished}.

\section{Conclusion}
The matter radius and the $E1$ response of S-wave 2$n$ halo nuclei were studied. 
We put constraints on the $(E_B,E_{nc})$ parameter space using the value of the ${}^{22}$C matter radius.
The calculated $B(E1)$ spectrum of $^{11}\mathrm{Li}$ agrees with the experimental result within our theoretical uncertainty. Our $^{22}\mathrm{C}$ 
result can be tested once the experimental data is available. Further improvements can be made by rigorously calculating the higher-order terms in the EFT expansion and 
by including higher partial waves. 

\smallskip

We thank our collaborators Chen Ji, Hans-Werner Hammer and Philipp Hagen. This work was supported by the US Department of Energy under grant DE-FG02-93ER40756. 
BA is grateful to the organizers of the conference for the opportunity to present this work and to UT for sponsoring his attendance.

%

\newpage


\begin{thebibliography}{}
%
%
\bibitem{Tanaka:2010zza} 
  K.~Tanaka et al.,
  Phys.\ Rev.\ Lett.\  {\bf 104}, 062701 (2010).
     
\bibitem{Bertulani:2002sz} 
  C.~A.~Bertulani, H.-W.~Hammer and U.~van Kolck,
  Nucl.\ Phys.\ A {\bf 712}, 37 (2002).

  \bibitem{Bedaque:2003wa} 
  P.~F.~Bedaque, H.-W.~Hammer and U.~van Kolck,
  Phys.\ Lett.\ B {\bf 569}, 159 (2003). 
  
\bibitem{Kaplan:1998we} 
  D.~B.~Kaplan, M.~J.~Savage and M.~B.~Wise,
  Nucl.\ Phys.\ B {\bf 534}, 329 (1998).
  
  \bibitem{Bedaque:1998kg} 
  P.~F.~Bedaque, H.-W.~Hammer and U.~van Kolck,
  Phys.\ Rev.\ Lett.\  {\bf 82}, 463 (1999).
  
    \bibitem{Kobayashi:2011mm} 
  N.~Kobayashi et al.,
  Phys.\ Rev.\ C {\bf 86}, 054604 (2012). 
  
  
\bibitem{Acharya:2013aea} 
  B.~Acharya, C.~Ji and D.~R.~Phillips,
  Phys.\ Lett.\ B {\bf 723}, 196 (2013).


  
  \bibitem{Langevin}
M.~Langevin et al.,
Phys. Lett. B, {\bf150}, 71 (1985). 
  
  \bibitem{Mosby:2013bix} 
  S.~Mosby et al.,
  Nucl.\ Phys.\ A {\bf 909}, 69 (2013).

\bibitem{Nakamura:2006zz} 
  T.~Nakamura et al.,
  Phys.\ Rev.\ Lett.\  {\bf 96}, 252502 (2006).

  
  
  \bibitem{Acharya:tobepublished} 
  B.~Acharya , P.~Hagen, H.-W. Hammer and D.~R.~Phillips, 
  In preparation.
  
  
  
   \bibitem{Yamashita:2011cb} 
  M.~T.~Yamashita et al.,
  Phys.\ Lett.\ B {\bf 697}, 90 (2011)
  [Erratum-ibid.\ B {\bf 715}, 282 (2012)].
  
  \bibitem{Fortune:2012zzb} 
  H.~T.~Fortune and R.~Sherr,
  Phys.\ Rev.\ C {\bf 85}, 027303 (2012). 
    
  \bibitem{Gaudefroy:2012qe} 
  L.~Gaudefroy  et al.,
  Phys.\ Rev.\ Lett.\  {\bf 109}, 202503 (2012).

  
\bibitem{Smith:2008zh} 
  M.~Smith et al.,
  Phys.\ Rev.\ Lett.\  {\bf 101}, 202501 (2008).
  
 \bibitem{NNDC}
National Nuclear Data Center, BNL,
Chart of Nuclides (2013), {\it http://www.nndc.bnl.gov/chart/}~.

\bibitem{Ershov:2012fy} 
  S.~N.~Ershov, J.~S.~Vaagen and M.~V.~Zhukov,
  Phys.\ Rev.\ C {\bf 86}, 034331 (2012).

    
\bibitem{Nakamura:2013conference} 
  T.~Nakamura,
  J.\ Phys.\ Conf.\ Ser.\  {\bf 445}, 012033 (2013).
  
\end{thebibliography}
%
%

\end{document}